# POWTEX visits POWGEN

Andreas Houben[a], Philipp Jacobs[a], Yannick Meinerzhagen[a], Noah Nachtigall[a], and Richard Dronskowski[a*]

[a] Chair of Solid-State and Quantum Chemistry, Institute of Inorganic Chemistry, RWTH Aachen University, D-52056 Aachen, Germany

Correspondence email: drons@HAL9000.ac.rwth-aachen.de

**Abstract**

The high-intensity time-of-flight neutron diffractometer POWTEX for powder and texture analysis is currently being built to be operated in the eastern guide hall of the research reactor FRM II near Munich. Having heavily faced the world-wide $^3$He crisis in 2009, we promptly initiated the development of $^3$He-free detector alternatives that are tailor-made for the requirements of large-area diffractometers. Herein, we report about the 2017 enterprise to operate one building unit of the final POWTEX detector at the neutron powder diffractometer POWGEN at the Spallation Neutron Source of the Oak Ridge National Lab, USA. As a result, we present the first angular- and wavelength-dependent data of the POWTEX detector, current data-reduction steps as implemented in Mantid [1] and also multi-dimensional refinement results of two samples (diamond and BaZn(NCN)$_2$) by use of the GSAS-II software suite.

**Introduction**

The POWTEX instrument is a pulsed high-intensity time-of-flight diffractometer at a continuous reactor source. Its design utilizes several novel concepts and is suited for small samples. [2, 3] The jalousie-design large-area detector-system was tailor-made to suffice the resolution requirements and instrument characteristics of POWTEX and has been described in detail already [4]. The almost blind-spot free alignment of the cylindrical shape around the cubic sample is realized by two detector building types following the same detection principle. Manufacturing the major instrument components of POWTEX, including the surface detectors, i.e., those at $2\theta \approx 45 - 135°$, has been accomplished. Due to several delays which are not further commented herein, we couldn't and still cannot operate POWTEX at the FRM II so far. Since we are actively developing multi-dimensional data-treatment and -analysis methods for angular- and wavelength-dependent diffraction data [5, 6], there was an obvious need to find a proper place to test our brand-new, 240 cm long detectors under realistic (i.e., POWTEX-like) diffraction conditions. Hence, the POWGEN instrument at SNS, ORNL, was an ideal choice for several reasons. First, POWGEN is a latest-generation time-of-flight diffractometer with a huge, large-area detector-coverage. [7] Second, prior to the recent upgrade program, there was sufficient space to install one POWTEX detector unit centered at $2\theta = 90°$ at the POWTEX-sample-detector distance of 80 cm. Third, we knew the POWGEN instrument quite well from user beam-times and especially also from our methodical work [8], for which comparing the two instrument designs seems very promising.

It took one year from presenting our ideas to Ashfia Huq, the leading instrument-scientist at those times, until the scheduled experiment in Nov. 2017, carefully and narrowly timed between the end of the user-program cycle and POWGEN's own upgrade-program thereafter, finally closing the needed space for the POWTEX detector. Many documentations, agreements, and meetings later the detector was shipped to the ORNL, where it finally arrived with several 50 $g$ shock-watches having triggered and the detector being most possibly damaged. In a way, this very shock translated into the instrument developers, too. Because of the unique chance to measure POWTEX@POWGEN, we tried to ship a replacement detector but this would not have arrived on time. After preliminarily investigating the damage, several anode-wires turned out as being broken, at least one corner of the detector-hood was dented, and a delicate rattling sound could be heard from within when the detector was shaken gently. This being a fine piece of German engineering, we still took the opportunity to use the test-beam-time such as to characterize the damage and possibly learn about the detector as well.

1. **Beam-time and raw data conversion**

In contrast to earlier tests targeting detector-physical properties [4], this beam-time aimed to collect diffraction data with a setup as close as possible to the later POWTEX setup (see Figure 1). While the entire primary instrument (neutron guides, choppers, slits, etc.) was different from POWTEX, i.e., identical to the POWGEN instrument, the secondary instrument completely consists of POWTEX components (except for the neutron windows in the POWGEN sample vessel). That means that, in addition to the detector modules themselves, also the analog and digital electronics, the clock-sync-distribution, the high-voltage supply, the gas handling system, the detector firmware, etc. were all used as planned for POWTEX. The T0-chopper signal of POWGEN was fed into our detector-clock to sync with the SNS pulse generation.

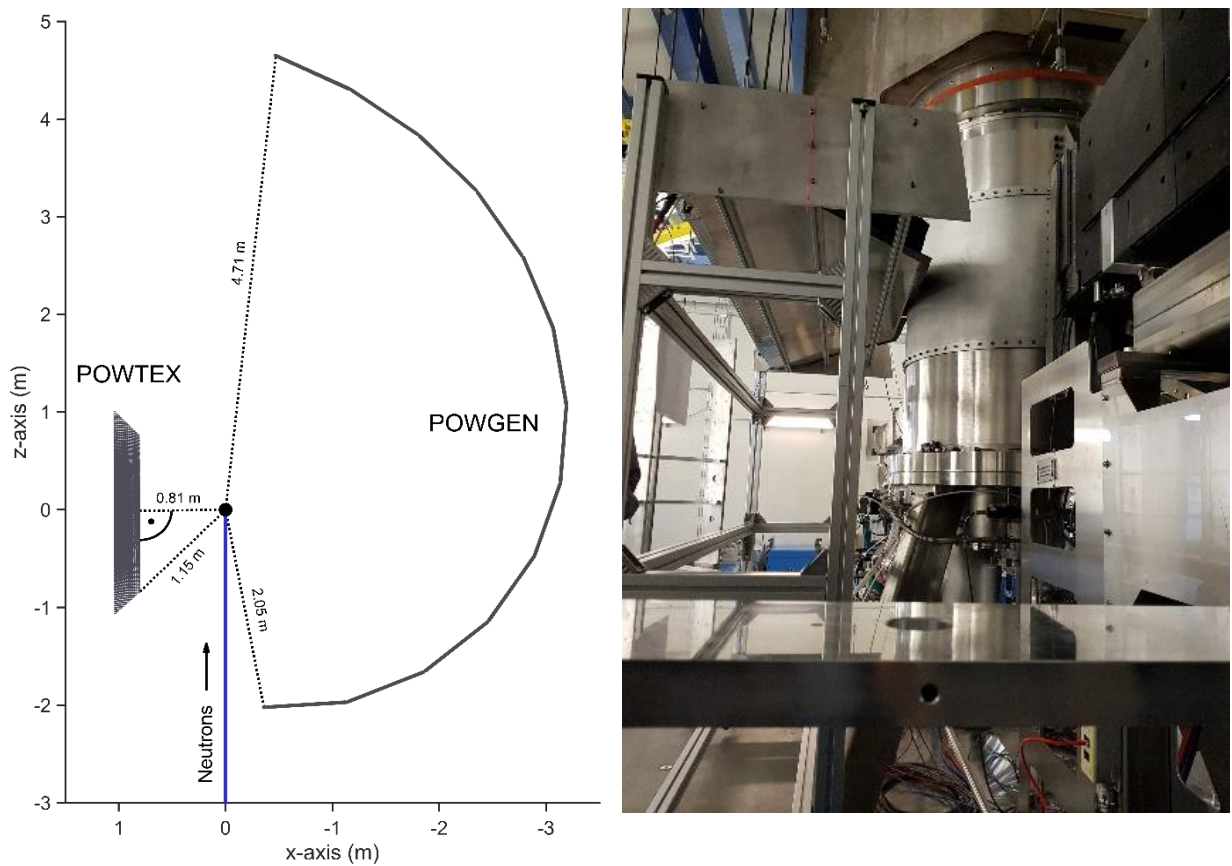

*Figure 1: Left: schematic drawing showing one of the POWTEX cylinder-surface detector-modules (left part) mounted in the POWTEX detector-geometry: aligned orthogonal to 2θ = 90° in the horizontal scattering-plane (paper sheet) with 81 cm (instead of ideal 80 cm) distance to the sample at the center of the coordinate system. The drawing results from the instrument definition file (IDF) as used with Mantid. The much larger POWGEN detector system is shown on the opposite hemisphere. Right: picture of the detector (plus mockup and shielding) touching the POWGEN detector vessel, i.e., the neutron window, as close as possible to best match the 80 cm sample-detector-distance.*

In the main test phase, the setup was continuously running over six days without any failure, constantly collecting neutron events. As a sad consequence of the transport damage, two and a half of the total eight subunits had to be disabled completely because their anode wires were broken producing a short circuit. It may well be the case that the remaining six subunits are the only 50 *g* shock-proven neutron-detectors world-wide.

Next to the standard samples used to also calibrate POWGEN, e.g., highly crystalline diamond powder, a vanadium rod, a NAC sample ($Na_2Ca_3Al_2F_{14}$, space group $I2_13$, *a* = 10.251 Å) and empty POWGEN sample holders (background), three real-world user beam-times of "friendly scientists" were simultaneously measured at the POWGEN instrument and on the POWTEX detector. Herein, we will only refer to the data sets for diamond, vanadium and background, as well as those of BaZn(NCN)$_2$. Vanadium primarily scatters incoherently and isotropic in space, allowing for a correction of the detector efficiency per detector-voxel, i.e., a pixel of a volume detector. Furthermore, geometrical effects such as the Lorentz effect [9] and the source spectrum, i.e., the TOF-dependent intensity distribution on the path from the source to the sample, are intrinsically accounted for. The background measurement is subtracted to remove "noise" from other scattering sources. The diamond sample with its very sharp reflections is used to correct the detector voxel-positions (or TOF-offsets) within the overall instrument geometry and compared to the ideal instrument definition.

Since the POWTEX detector was continuously running, the continuous raw-data files were first cut into pieces according to the time stamps of the NeXus [10] files generated for POWGEN using a routine supplied by Günter Kemmerling, JCNS, to yield a set of data files correlated with the parallel measurement runs at POWGEN. Afterwards, those raw files were converted to NeXus files by a conversion routine supplied by Gerd Modzel, CDT GmbH, later JCNS. During the conversion, the meta data (sample name, chopper , instrument settings, etc.) which were identical for both experiments were copied from the POWGEN NeXus file, while the event data were converted from raw POWTEX data collected at a given time stamp. All further analysis and data treatment routines described in the following have been designed to work with the NeXus format, since it will be the future event file format for POWTEX@FRM II.

For a future machine such as POWTEX, a neutron event in a time-of-flight experiment using a volume detector consists of three spatial coordinates and a time channel. However, in NeXus files only voxel IDs as written by the firmware are stored while the mapping to (diffraction) coordinates needs to be done separately and as needed for the data treatment. In Mantid this is achieved by an instrument definition file (IDF), which was created (and iterated) according to the analysis of the transport damage. Simply speaking, it holds each voxel's center of gravity position and its hexahedron-like shape connected with a unique detector ID. In the design of the POWTEX jalousie detector, each voxel has its slightly different (individual) shape.

## 2. Analysis of transport damage

The installation and alignment of each instrument component involves small deviations from the ideal positioning as defined in the IDF, especially in a test setup and even more if the detector was damaged before. Therefore, there are differences $\Delta d_{\text{offset}}$ between the theoretical and the actual voxel position of the detector and, consequently, also discrepancies regarding the interplanar distances $d$ calculated therefrom and which finally shall be measured. $\Delta d_{\text{offset}}$ is determined routinely at neutron diffractometers, e.g., at the beginning of each measurement cycle using a standard sample. For POWGEN, this is usually a polycrystalline, powdered diamond sample. For technical reasons, the differences are treated by the introduction of a correction factor $difC$:

$$difC = \frac{TOF_{exp}}{d_{theo}}$$

All possible effects causing a spatial dislocation (alignment effects, detector damage, etc.) of a voxel need to be corrected by this factor prior to further evaluation of usual diffraction data sets.

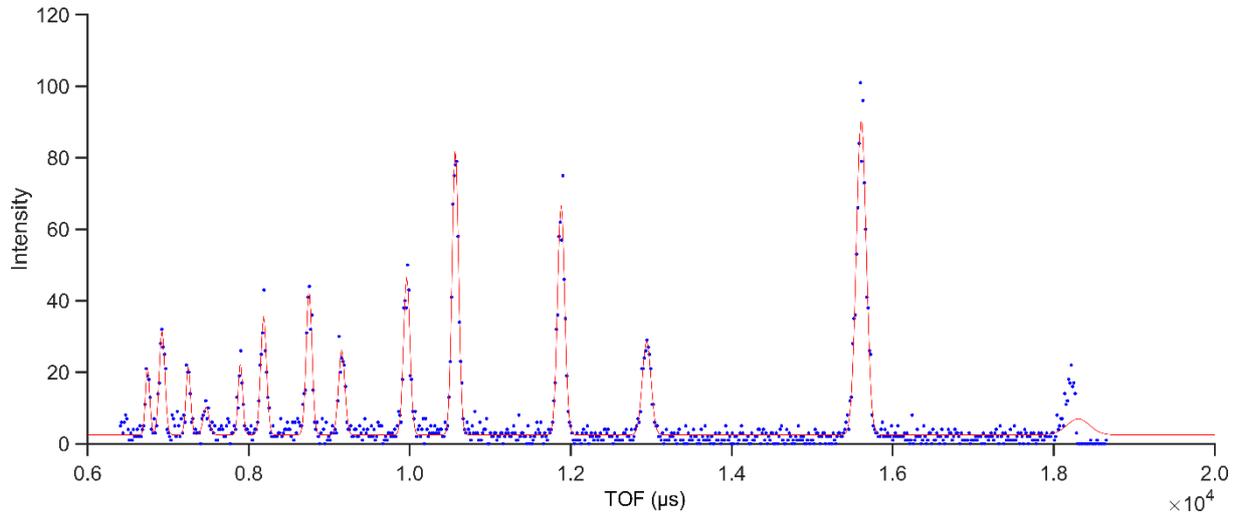

*Figure 2: Example diffractogram of the TOF values as measured similarly for each voxel. The experimental data points are shown as blue dots and the fitted function as a red line.*

The first step in correcting the voxel positions is to determine the current time-of-flight value for each peak on each voxel. The NeXus files contain the TOF values for each neutron event in a voxel with its respective detector ID. Each of those diffractograms per each voxel is fitted using a constant peak-function type and one peak-function per expected reflection position. Each peak-function is defined by the theoretical *d*-value, by the peak-shape parameters and, most importantly, by the offset value to be determined. With the exception of this very offset, the other peak-shape parameters should be known in advance, as the offset value must be fitted to the experimental data based on a known peak-shape and position.

Hereby, it is of crucial significance that the description of the experimental data is consistent throughout the entire process of data treatment. For example, it is quite common to use either pseudo-Voigt functions or pV-functions convoluted with back-to-back exponentials [11] (pV-b2b) for the correction. Since the SNS is a short pulse source, the reflections show the known asymmetries, and it is advisable to use the back-to-back exponentials for the correction in this case, as is done for the POWGEN instrument [12, 13]. Unfortunately, for the many voxels and reflections this is computationally demanding, especially when using the back-to-back exponentials. Herein, only the Gaussian part was used for the correction (equiv. to a pV-mixing-parameter $\eta = 0$). Since this choice has no effect on the $d$-offset, it is a rather safe step to do. For an additional time saving, only the Gaussian functions were used in a first run to calculate starting values $S, H$, (see below) for a second fit using pV-b2b-functions. The equations for the fitting functions are presented below:

$$I_{gauss,hkl} = S \cdot \frac{2}{H} \cdot \sqrt{\frac{\log(2)}{\pi}} \cdot \exp\left(-4 \cdot \frac{\log(2)}{H^2} \cdot \Delta d_{offset}^2\right)$$

$$I_{gauss+b2b,hkl} = S \cdot \left(N\left(\exp\left(\frac{1}{2} \cdot \alpha + \cdot (\alpha \cdot \sigma^2 + 2\Delta d_{offset})\right) \cdot \text{erfc}\left(\frac{\alpha \cdot \sigma^2 + \Delta d_{offset}}{\sqrt{2 \cdot \sigma^2}}\right)\right.\right.$$

$$\left.\left. + \exp\left(\frac{1}{2} \cdot \beta \cdot (\beta \cdot \sigma^2 - 2\Delta d_{offset})\right) \cdot \text{erfc}\left(\frac{\beta \cdot \sigma^2 - \Delta d_{offset}}{\sqrt{2 \cdot \sigma^2}}\right)\right)\right)$$

$$\sigma = \sqrt{H^2/(8 \cdot \log 2)}$$

$$N = (\alpha \cdot \beta)/(2\alpha + 2\beta)$$

$$\Delta d_{offset} = t/difC_{offset} - d_{hkl}$$

with $S$ = scale, $H$ = full-width-at-half-maximum, $\Delta d_{offset}$ = offset in Å, $t$ = time-of-flight, $difC_{offset}$ = offset in $\frac{\mu s}{Å}$ and $d_{hkl}$ = theoretical $d$-value of the peak. The fit results in a data set with $difC_{offset}$ values for each detector ID (voxel). From the offset and the theoretical peak positions an experimental TOF value can be calculated $\text{TOF}_{exp} = d_{hkl} \cdot difC$. With $\text{TOF}_{exp}$, $d_{theo}$ and the uncorrected $x$, $y$ and $z$ coordinates of each voxel, new $x'$, $y'$ and $z'$ coordinates can be calculated:

$$d_{exp} = \frac{10^{10}\frac{Å}{m} \cdot \text{TOF}_{exp} \cdot h_p}{2 \cdot 10^6 \frac{\mu s}{s} \cdot m_N \cdot \left(\sqrt{x'^2 + y'^2 + z'^2} + 60\right) \cdot \sin\left(\frac{1}{2} \cdot \arccos\left(\frac{z'}{\sqrt{x'^2 + y'^2 + z'^2}}\right)\right)}$$

$$\begin{pmatrix} x' \\ y' \\ z' \end{pmatrix} = \text{rotz}(RZ) \cdot \left(\text{roty}(RX) \cdot \left(\begin{pmatrix} x \\ y \\ z \end{pmatrix} - \begin{pmatrix} x_0 \\ y_0 \\ z_0 \end{pmatrix}\right)\right) + \begin{pmatrix} DX \\ DY \\ DZ \end{pmatrix} + \begin{pmatrix} x_0 \\ y_0 \\ z_0 \end{pmatrix}$$

To achieve this, the rotation and translation parameters $RX, RY, RZ, DX, DY, DZ$ are fitted until the experimental $d_{exp}$ position matches the theoretical $d_{theo}$ position.

Using this set of parameters, the global detector geometry could be determined, and a new instrument definition file (IDF) was created for the new voxel coordinates.

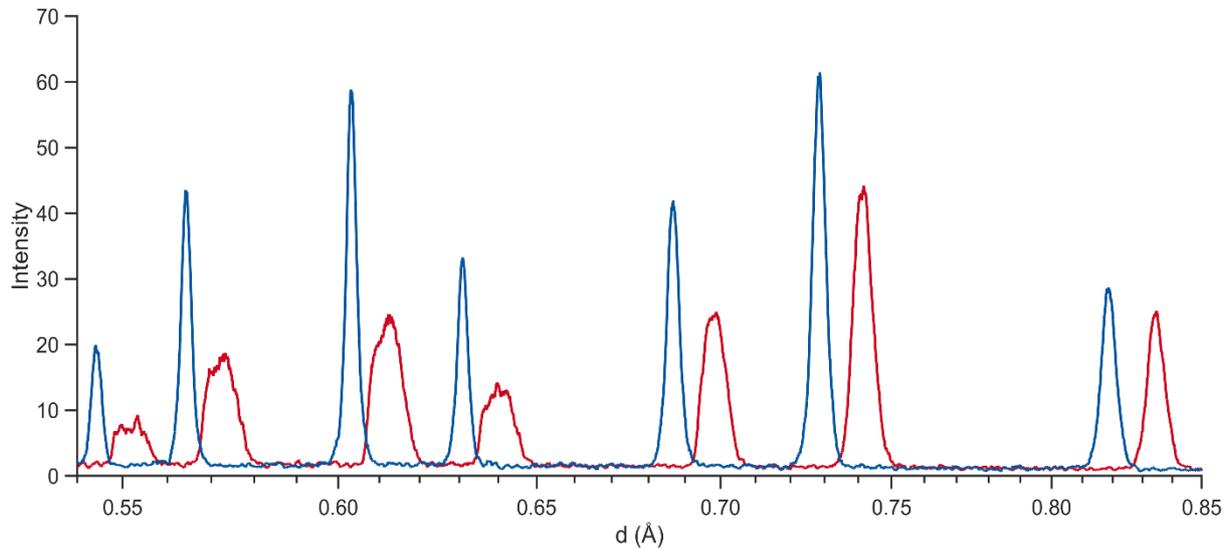

*Figure 3: One-dimensional diffractogram for the diamond sample measured at POWTEX@POWGEN, in red for the uncorrected and in blue for the corrected detector alignment also describing the detector damage.*

Figure 3 shows the difference between the uncorrected (red) and corrected (blue) detector alignment in a one-dimensional diffractogram ($I$ vs. $d$). Two effects can be observed: there is a general shift in the $d$-position of the peaks as well as a broadening of the peak-shape in $d$. The discrepancies in the peak-shape can be better illustrated in a two-dimensional drawing.

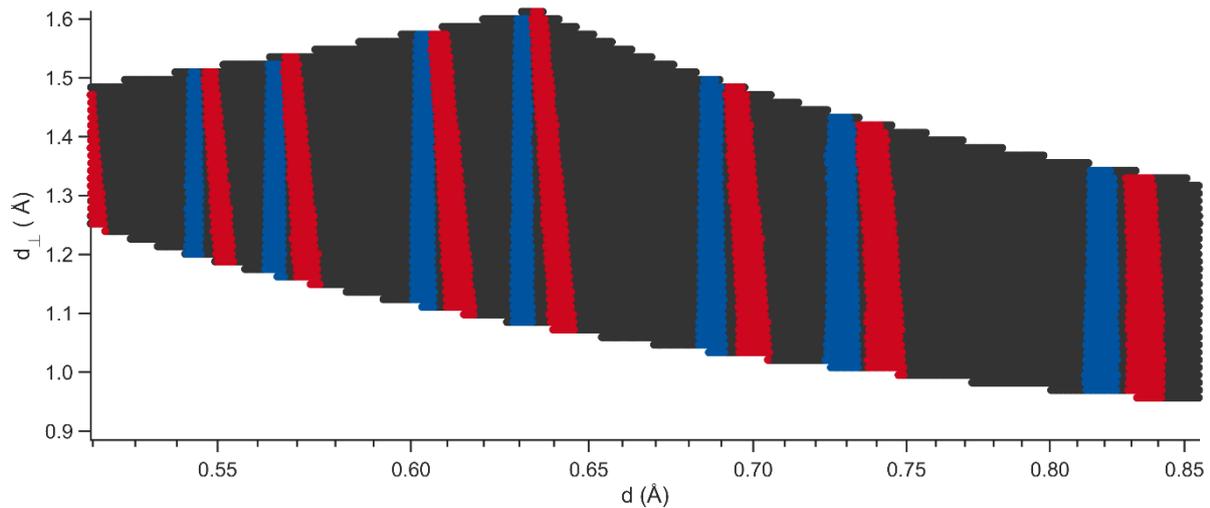

*Figure 4: Two-dimensional diffractogram for the diamond POWTEX@POWGEN data sets with an uncorrected (red) and corrected (blue) detector alignment also handling the effects of the detector damage.*

Figure 4 shows the same data set as in Figure 3 as a two-dimensional $d_\perp$–$d$ scheme reduced to only the information about the position of the peaks and thus neglecting intensities. For the introduction of $d_\perp$ please see [5]. This reveals that the peak-shape does not necessarily change, but that the $d$-position is shifted to smaller $d$-values over the course of $d_\perp$. As a result, the red peaks in Figure 4 are skewed and, accordingly, the red reflections in Figure 3 are broadened and misshaped. *With applied correction (blue), the peaks are aligned along $d_\perp$ at constant $d$-values as required by the definition and also show a proper one-dimensional diffractogram* (Figure 3). By having thus acquired both ingredients, proper NeXus files like for the later POWTEX and a corrected IDF file properly describing the detector alignment and the detector damage, it is then possible to investigate the Mantid routines for handling those data sets.

## 3. Data reduction in Mantid

The data reduction of the raw event data has been accomplished using the Mantid software [1]. As there was no existing algorithm for reducing angular- and wavelength dependent data, the new workflow algorithm *PowderReduceP2D* was implemented into the Mantid software. *PowderReduceP2D* (for technical details see [14]) reduces raw event data following three basic steps (see Figure 5). Those steps include calibration, binning, and correction of the raw event data.

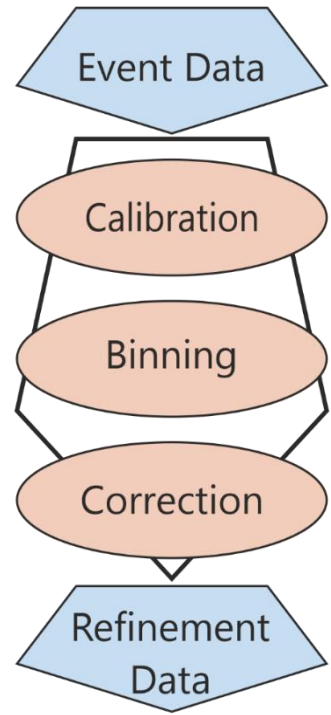

*Figure 5: The three basic steps of data reduction applied by the workflow algorithm PowderReduceP2D to convert raw event data to refineable p2d data.*

For the reduction of raw sample data, additionally two "standard measurements" should be supplied as well, namely, a measurement of an empty sample holder and of a vanadium sample. The empty measurement serves to reduce background noise, while the vanadium measurement allows to normalize differences in detector voxel efficiency and other effects (see above). During the reduction, all three sets of raw data are basically treated the same. Only the vanadium set needs some additional corrections which will be explained at the relevant steps.

The first step, calibration, applies several already existing Mantid algorithms for inspecting the neutron pulse, calibrating the detector positions, and masking the detectors if needed. The algorithm *FilterBadPulses* is used to remove events that occurred while the proton charge is below a certain threshold. Additionally, the algorithm *RemovePromptPulse* (as one can guess from its name) removes the prompt pulse from the measuring data. The detector geometry is computed using the algorithm *FindDetectorsPar* and further calibrated by the algorithm *AlignDetectors*.

*AlignDetectors* uses a calibration file that was once generated using the routine *CalibrateRectangularDetectors* (until Mantid 6.2) with the diamond measurement and a list of reflection positions as input to again correct detector positions. In contrast to the calibration procedure described above, this function is capable of handling rather tiny displacements of individual voxels. As described in the section above, it was necessary to first adjust the voxel positions in the IDF file in order to describe the alignment and detector-damage effects to the detector as a whole and afterwards realign the individual voxel positions to generate the needed calibration files for the following routines. Hereby, also certain voxels showing too large offsets or being malfunctioning where masked using the algorithm *MaskDetectors* (see Figure 6).

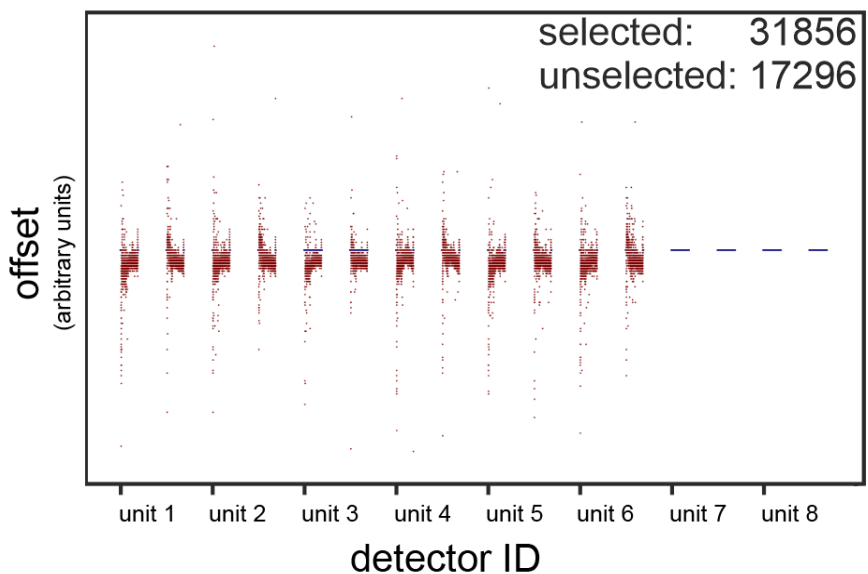

*Figure 6: The diagram shows the offsets assigned to each voxel as identified by their detector id. Red voxels were selected for further data treatment while blue voxels where masked for different reasons. Each of the eight modules per building unit consist of two sub-units. Therefore, the four subunits, i.e., the two modules with highest detector IDs, were masked because their anode wires were broken. Similarly, in the third module half of the detector volume (in depth) had to be shut down.*

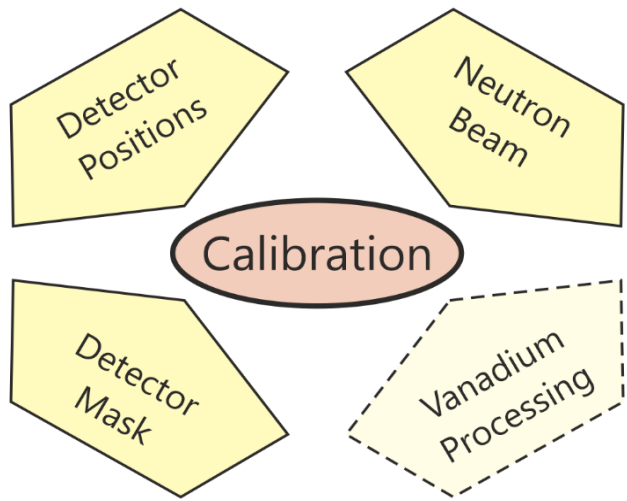

*Figure 7: Calibration of the raw event data includes examining the neutron beam, loading and correcting the detector positions and applying a mask to the detectors if needed. Vanadium data are further processed to remove influences from cylindrical absorption.*

In case of reducing raw vanadium event data, further processing is done at this point. The algorithm *CylinderAbsorption* is applied to calculate correction factors for attenuation caused by absorption and scattering inside a cylindrical sample.

The second step for data reduction is the binning of the raw event data in two dimensions. The algorithm *Bin2DPowderDiffraction* bins the event data either in classical linear/logarithmic bins or applies the recently developed edge binning [15]. If edge binning is used, a file containing bin limits has to be supplied. In the case of linear/logarithmic binning, it is sufficient to supply a bin width (conventions as also used for other routines: positive bin width for linear binning, negative bin width for logarithmic binning). In case of binned vanadium data further processing using the algorithm *StripVanadiumPeaks* is necessary to remove small vanadium reflections. The algorithm *FFTSmooth* is then used to compensate noise utilizing the Fourier transform to filter higher frequencies.

The third step for data reduction is the correction of the measuring data. The correction follows equation (*1*) for each of the data bins:

$$y_{sample,corrected} = \frac{y_{sample} - y_{empty}}{y_{vanadium}} \quad (1)$$

First the empty measurement data is bin-wise subtracted from the sample measurement data to remove any background noise created by the sample holder. Second the resulting data are divided by the vanadium measurement data to compensate any differences of detector voxel efficiencies, etc. Finally, a fully reduced two-dimensional sample-data is created as text-file with suffix ".*p2d*" by calling the *SaveP2D* function of Mantid. A *p2d*-file contains information about the used instrument and the applied binning in the header as well as data-columns for $2\theta$, $\lambda$, $d$, $d_\perp$, binned intensity and its standard deviation. Additionally, the function allows to select the range of data to write into the *p2d*-file both in $d/d_\perp$ and $2\theta/\lambda$ space. This data file can be used as input data for a software like GSAS-2, which was modified to do a structure refinement with multi-dimensional data according to the Rietveld method (see section 4). Additionally, an instrument parameter file is required for the refinement process. Next to the instrument setup, herein a set of instrument-specific parameters is given, for example, to analytically calculate the peak-width and -shape during the refinement. These parameters are usually determined according to a well-known setup by measuring a reference sample, in this case diamond, and they remain unchanged in subsequent refinements of user data. Herein, an extended parametrization compared to the one described in [6] for the POWGEN instrument was used. The parametrization details clearly go beyond the scope of this article and will be reported separately later.

## 4. Multi-dimensional Refinement

For refining the created *p2d*-data-files, GSAS-II was modified to treat multi-dimensional neutron time-of-flight data. In this study two exemplary refinements of diamond and BaZn(NCN)$_2$ are shown.

The refinement of the diamond (for structural information please see [16]) data was accomplished in four steps:

1. Background refinement using a two-dimensional Chebyshev formula [17] with 15 parameters in both dimensions each. The procedure is very similar to the conventional treatment.
2. Refinement of the cell parameters.

3. Refinement of isotropic thermal parameters.
4. Refinement of instrumental parameters for further use, because of diamond being a standard sample.

The result of the multi-dimensional refinement is depicted in Figure 8, in particular showing in a) the observed pattern, in b) the calculated pattern, in c) the difference pattern and in d) a one-dimensional "standard" plot of the multi-dimensional refinement. For comparison, e) depicts a conventional, one-dimensional refinement of the same but conventionally reduced measuring data. The color-scale shows normalized intensity values because the sample measurement was divided by a vanadium measurement during data reduction. It is immediately obvious that the calculated pattern is a particularly good representation of the measured data according to the structural description of the diamond sample. There are only minor deviations visible in plot c) located at the peak positions and correlating with the peak intensity. The corresponding $R_{Bragg}$ values as well as cell parameters are given in Table 1. While the only carbon atom position on the special Wyckoff site 8$a$ is fixed, the isotropic thermal displacement parameters arrived to $U_{iso}$ = 0.0005(1) Å² in the conventional and $U_{iso}$ = 0.00171(1) Å² in the multi-dimensional refinement.

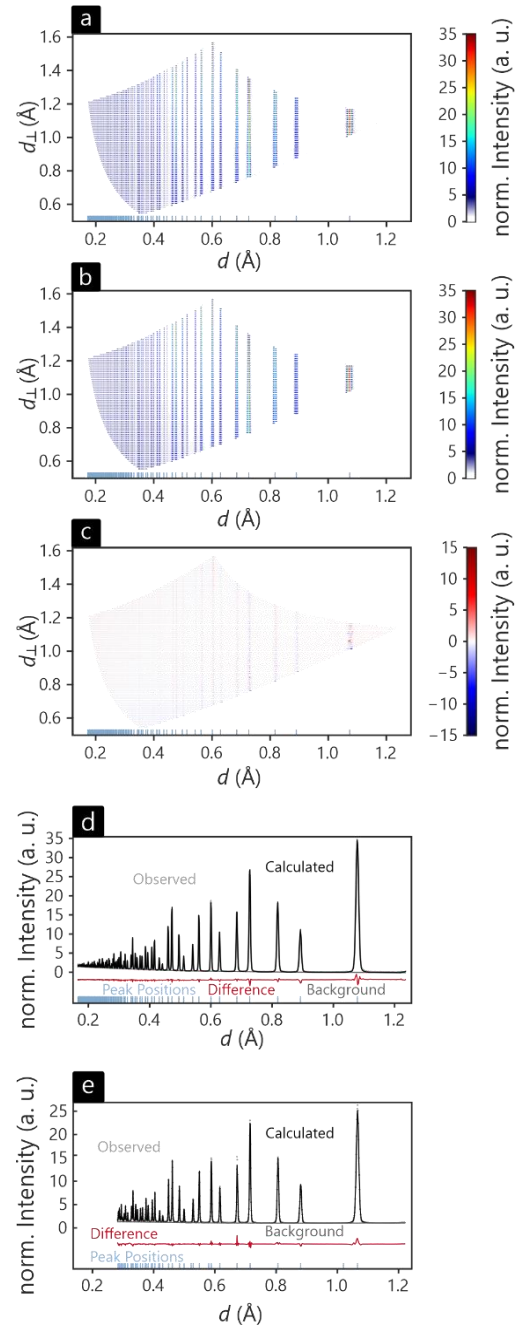

Figure 8: Result of the multi-dimensional Rietveld refinement for a neutron time-of-flight measurement of diamond. a) observed pattern, b) calculated pattern, c) difference pattern, d) 1D plot of multi-dimensional refinement, e) conventional, one-dimensional refinement. Blue lines indicate peak positions with intensity greater zero. The color scale shows normalized intensity in plots a) – c).

To exemplify that the multi-dimensional Rietveld refinement can be easily applied not only to highly crystalline diamond samples (with which also all correction work was done) a data set of a real-world (and timely) user sample, BaZn(NCN)$_2$, was analyzed as a second example. This phase is a ternary carbodiimide (a nitrogen-based pseudooxide with a complex NCN$^{2-}$ anion) with tetrahedrally coordinated Zn$^{2+}$ and eightfold Ba$^{2+}$ coordination, crystallizing in *Pbca* with $a$ = 11.934, $b$ = 11.927, and $c$ = 6.845 Å from powder XRD [18]. Here, the data range in $d$ was restricted to 0.65–1.08 Å due to challenges in calculating a multi-dimensional background description. The refinement of the BaZn(NCN)$_2$ data was accomplished in three steps:

1. Background refinement using Chebyshev formula [17] with 15 parameters in both dimensions each.
2. Refinement of cell parameters.
3. Refinement of isotropic thermal parameters.
4. Refinement of atomic positions.

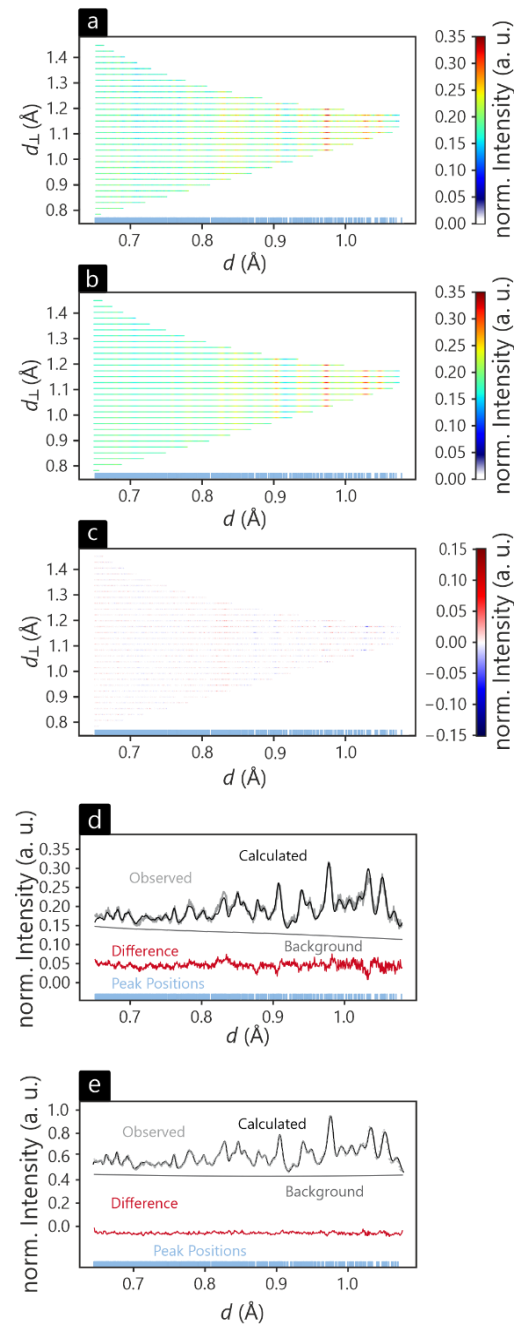

*Figure 9: Result of the multi-dimensional Rietveld refinement for a neutron time-of-flight measurement of BaZn(NCN)$_2$. a) observed pattern, b) calculated pattern, c) difference pattern, d) 1D plot of multidimensional refinement, e) conventional, one-dimensional refinement result. Blue lines indicate peak positions with intensity greater zero. The color scale shows normalized intensity in plots a) – c).*

Instead of refining the instrument parameters, we re-used the results from the diamond refinement.

The result of the refinement is displayed in Figure 9 by showing in a) the observed pattern, in b) the calculated pattern, in c) the difference pattern and in d) a one-dimensional "standard" plot of the multi-dimensional refinement. Again, part e) offers a conventional, one-dimensional refinement of the likewise conventionally reduced measured data for comparison. The color scale refers to a normalized intensity because the sample measurement was renormalized by division of a vanadium measurement. Plots a) and b) differ greatly from the according plots in Figure 8 due to the low sample crystallinity and many peaks overlapping in the available *d*-range. With the naked eye it is almost impossible to differentiate between single peaks (mirroring the proximity of the *a* and *b* lattice parameters) or to describe the background properly. Nonetheless, plots c) and d) evidence that the refined structural parameters describe the observed data very well. In particular, plot d) makes it clear that the observed and calculated patterns match with the difference pattern showing primarily noise. The seemingly larger noise in the difference-pattern of part d) as compared to the one of part e) goes back to the smaller intensity range caused by differences in the normalization.

The corresponding $R_{Bragg}$ values as well as cell parameters are given in Table 1. More detailed information on spatial and isotropic thermal displacement parameters are provided in Table 2.

Table 1: Refinement results for the conventional, one-dimensional (1D) and the multi-dimensional refinements of diamond and BaZn(NCN)$_2$ with corresponding $R_{Bragg}$ values (2D). In brackets, the tripled estimated standard deviations are shown for the last digit.

|  | Diamond; 1D | Diamond; 2D | BaZn(NCN)$_2$; 1D | BaZn(NCN)$_2$; 2D |
|---|---|---|---|---|
| $a$ (Å) | 3.57162(8) | 3.56663(3) | 11.953(4) | 12.025(4) |
| $b$ (Å) | 3.57162(8) | 3.56663(3) | 11.945(4) | 11.987(5) |
| $c$ (Å) | 3.57162(8) | 3.56663(3) | 6.855(2) | 6.871(2) |
| $V$ (Å$^3$) | 45.561(4) | 45.371(1) | 978.86(4) | 990.53(4) |
| $R_{Bragg}$ | 3.36% | 7.49% | 1.35% | 7.74% |

Table 2: Atom positions in fractional coordinates and isotropic thermal displacement parameters for BaZn(NCN)$_2$. The upper values are attributed to the one-dimensional refinement, the lower values are attributed to the multi-dimensional refinement.

|  | $x$ | $y$ | $z$ | $U_{iso}$ (Å$^2$) |
|---|---|---|---|---|
| Ba | 0.846(3) | 0.869(3) | 0.029(4) | −0.001(5) |
|  | 0.846(2) | 0.872(2) | 0.030(4) | 0.000(4) |
| Zn | 0.920(3) | 0.122(3) | 0.283(5) | 0.006(5) |
|  | 0.920(3) | 0.125(4) | 0.283(5) | 0.010(4) |
| C1 | 0.368(3) | 0.163(3) | 0.601(4) | 0.005(4) |
|  | 0.366(2) | 0.161(3) | 0.609(4) | 0.008(4) |
| C2 | 0.613(3) | 0.424(3) | 0.522(5) | 0.006(4) |
|  | 0.613(3) | 0.426(3) | 0.523(5) | 0.012(4) |
| N1 | 0.653(3) | 0.494(2) | 0.416(4) | 0.017(4) |
|  | 0.656(3) | 0.496(2) | 0.416(4) | 0.018(4) |
| N2 | 0.419(2) | 0.095(2) | 0.512(3) | 0.011(4) |
|  | 0.418(2) | 0.095(2) | 0.511(3) | 0.014(4) |
| N3 | 0.317(3) | 0.243(2) | 0.678(4) | 0.015(5) |
|  | 0.316(3) | 0.241(2) | 0.676(4) | 0.017(5) |
| N4 | 0.579(3) | 0.348(2) | 0.627(3) | 0.013(5) |
|  | 0.580(2) | 0.349(2) | 0.630(3) | 0.015(4) |

## 5. Discussion

A closer look at the results of the diamond refinements from Table 1 yields that the one-dimensional and the multi-dimensional refinements lead to very similar results. Although the cell parameters and volumes do not arrive at exactly the same values, their precisions are alike. Accordingly, the precision of the isotropic thermal displacement parameters is alike as well. Also, the lattice parameter published in [16], $a$ = 3.5668 Å, coincides with the multi-dimensional refinement results.

The results for the BaZn(NCN)$_2$ sample point into the same direction; although the cell parameters and volumes in Table 1 are not quite identical, the precisions are alike, and this also related to the precisions of the isotropic thermal displacement parameters in Table 2. Additionally, the refined atomic positions for **all** atoms are identical within their tripled estimated standard deviations.

The $R_{Bragg}$ values given in Table 1 also indicate that those of the one-dimensional Rietveld refinements are smaller than those of the multi-dimensional refinement. This finding was already noted and discussed [5], and it may be a trivial consequence of the significantly larger number of data points used during the refinement; a similarly irritating size difference (but without physical meaning) between intensity-based $R$ values ($wR_2$) and structure-factor-based ones ($R_1$) was noted for single-crystal X-ray refinement when SHELXL-93 was released [19].

Lastly, Table 1 highlights an interesting observation: the carbodiimide phase crystallizing in *Pbca* is actually an intricate case for powder diffraction, simply due to the proximity of the *a* and *b* lattice parameters, differing by only 0.007 Å from powder XRD. Practically the same difference is found from the one-dimensional neutron refinement, 0.008 Å, mirroring the identical one-dimensional strategy. The two-dimensional neutron refinement, however, more clearly differentiates between *a* and *b*, by a five times larger 0.038 Å. One might think that this larger *a*/*b* difference goes back to the significantly larger amount of data points and/or a better profile model which the two-dimensional approach can provide. In addition to that, a preliminary analysis of the internal structural coordinates shows that the two-dimensional refinement yields more balanced interatomic distances but slightly sharper angles as regards the carbodiimide units. While the C–N bond lengths scatter between 1.18–1.25 Å in the 1D case, the 2D approach gives 1.21–1.25 Å with similar standard deviations around 0.013 Å. Also, the two N–C–N units are slightly less bent in 1D (173 and 175°) than in 2D (167 and 173°). As regards the tetrahedral coordination of divalent Zn, it is less regular in 1D (1.99–2.06 Å) than in 2D (2.02–2.06Å). Clearly, this needs a deeper investigation in the future.

These results demonstrate that multi-dimensional Rietveld refinement using a modified version of the GSAS-II software (and a partially defect detector) not only works but leads to results at least as precise as the conventional, one-dimensional Rietveld refinement. This further proves that the reduction of the raw measurement data as implemented in the Mantid software [1] is working adequately.

## 6. Conclusion and Outlook

In this present study we have shown the results of the detector test for the new neutron time-of-flight instrument POWTEX operated at the POWGEN instrument at SNS, Oak Ridge. After having overcome tremendous difficulties as a consequence of an unfortunate transport damage to the detector, we were still able to measure several data sets including diamond, vanadium, and BaZn(NCN)$_2$. Then, a new data reduction routine for multi-dimensional data sets was implemented into the Mantid software and used, successfully so, to reduce real-world multi-dimensional diffraction data using Mantid for the first time. These reduced multi-dimensional data sets were then refined, also for the first time, using a modified version of the GSAS-II software. Both refinements of diamond and BaZn(NCN)$_2$ went smoothly, and their results were compared to the conventional, one-dimensional approach of the same data sets. To conclude, the precision of the refined parameters was alike in one-dimensional and multi-dimensional refinements for both data sets. Additionally, the refined atomic positions of BaZn(NCN)$_2$ were identical within tripled standard deviations, with delicate differences as regards their internal structural coordinates. Hence, there is clear evidence that the multi-dimensional Rietveld refinement yields at least the same precision as the one-dimensional Rietveld refinement.

In addition, the proof-of-concept for multi-dimensional data-reduction using Matlab has been transferred to publicly available (open source) data reduction code in the widely used Mantid software. Also, multi-dimensional Rietveld refinement was transformed from the proof-of-concept utilizing Matlab [5] to a modified, not yet published version of the widely used software suite GSAS-II. A more technical description of the modified version shall be published in a separate article, once the code is publicly released. While all basic refinement steps are nicely working within GSAS-II, we are currently working to extend the refinement options towards those examples for which more demanding effects must be treated. For example, the present $BaZn(NCN)_2$ data set already posed a few challenges concerning the multi-dimensional background description, straightforwardly solved by providing the option of a manual background created from user-supplied base points. Also, the reason for the more pronounced difference between the *a* and *b* lattice parameters in the two-dimensional refinement of $BaZn(NCN)_2$ and its resulting internal coordinates as compared to the conventional refinement needs to be investigated carefully. Further, sample effects like stress/strain and hydrogen background will be investigated in the future; the latter incorporates a peculiar $d_\perp$-dependency which makes is uniquely treatable only within a multi-dimensional refinement.

## 7. Acknowledgements

Parts of this work are based upon data measured with the detector of POWTEX operated at the POWGEN instrument at the Spallation Neutron Source (Oak Ridge) and were thus sponsored by the Scientific User Facilities Division, Office of Basic Energy Sciences, U.S. Department of Energy. The authors want to especially thank Ashfia Huq and all people involved at the SNS for integrating our detector tests in the tight schedule of instrumental upgrade work of POWGEN including slight modifications of POWGEN; without their willingness, such great opportunity to do the critical experiment would have been impossible. Our warmest thanks also go to the many people at JCNS, RWTH Aachen, and CDT GmbH who were always helping us in those (sometimes) troublesome times. Eventually, we thank the unknown person who subjected the POWTEX detector, probably without bad intentions, to an unexpected 50 $g$ shock test, thereby proving that the POWTEX detector can safely take a devastating punch and still work beautifully after some tender care.

The financial support by the Federal Ministry of Research and Education (BMBF) for the POWTEX project (05K16PA2) is gratefully acknowledged.